\pdfoutput=1

\documentclass[11pt,twoside,a4paper,cmspaper,final,collab]{cms-tdr}
\def\svnVersion{421185P}\def\cmsCernNoTag{CERN-EP-2016-313}\def\cmsCernDate{\today}\def\cmsMessage{Published in the Journal of High Energy Physics as \href{http://dx.doi.org/10.1007/JHEP08(2017)046}{\doi{10.1007/JHEP08(2017)046.}}}

\begin{document}\cmsNoteHeader{FSQ-16-002}

\hyphenation{had-ron-i-za-tion}
\hyphenation{cal-or-i-me-ter}
\hyphenation{de-vices}
\RCS$HeadURL: svn+ssh://svn.cern.ch/reps/tdr2/papers/FSQ-16-002/trunk/FSQ-16-002.tex $
\RCS$Id: FSQ-16-002.tex 418233 2017-07-27 17:26:02Z alverson $

\cmsNoteHeader{FSQ-16-002}
\title{Measurement of the inclusive energy spectrum in the very forward direction in proton-proton collisions at $\sqrt{s} =13\TeV$}

\date{\today}

\abstract{The differential cross section for inclusive particle production as a function of energy in proton-proton collisions at a center-of-mass energy of 13\TeV is measured in the very forward region of the CMS detector.
The measurement is based on data collected with the CMS apparatus at the LHC, and corresponds to an integrated luminosity of 0.34\mubinv.
The energy is measured in the CASTOR calorimeter, which covers the pseudorapidity region $-6.6<\eta<-5.2$. The results are given as a function of the total energy deposited in CASTOR, as well as of its electromagnetic and hadronic components. The spectra are sensitive to the modeling of multiparton interactions in pp collisions, and provide new constraints for hadronic interaction models used in collider and in high energy cosmic ray physics.}

\hypersetup{%
pdfauthor={CMS Collaboration},%
pdftitle={Measurement of the inclusive energy spectrum in the very forward direction in proton-proton collisions at sqrt(s)=13 TeV},%
pdfsubject={CMS},%
pdfkeywords={CMS, CERN LHC, forward phase-space, CASTOR}}

\maketitle

\section{Introduction}
Particle production at very forward rapidities ($\abs{\eta}>5$) in high energy hadronic collisions receives contributions from multiparton interactions (MPI), initial- and final-state radiation (from the hardest parton scattering), fragmentation of the so-called ``beam remnants'', plus diffraction~\cite{N.Cartiglia:2015gve}.
The sum of all these particle production mechanisms, typically referred to as the ``underlying event''~\cite{Field:2002vt}, is modeled phenomenologically in hadronic Monte Carlo event generators with parameters tuned from the data~\cite{Sjostrand:1987su,Corke:2010yf,Skands:2014pea}.
A good understanding of forward particle production is important for a complete description of the final states in proton-proton (pp) interactions at colliders, as well as to accurately simulate extensive air showers induced in the earth atmosphere by very high energy cosmic rays~\cite{Ulrich:2010rg}.
In particular, forward charged hadron production has direct impact on the total number of air-shower muons at the ground, whose measurement shows unexplained excesses compared to model predictions~\cite{Aab:2014pza}.
The rapidities covered in the presented measurement ($-6.6 < \eta < -5.2$, still well separated from the beam rapidity $y \approx 9.5$) are mostly sensitive to the MPI activity~\cite{Chatrchyan:2013gfi}.

Previous studies of very forward particle production in pp collisions have been carried out at
center-of-mass energies of 0.9, 2.76, 7 and 8\TeV by CMS~\cite{Chatrchyan:2013gfi,Chatrchyan:2014qka}, at 7 and 8\TeV by TOTEM~\cite{Aspell:2012ux,Antchev:2014lez,Chatrchyan:2014qka},
and at 7\TeV by LHCf~\cite{Adriani:2011nf,Adriani:2015nwa}.
The present paper reports new measurements
of inclusive energy spectra at a center-of-mass energy of 13\TeV . The data are discussed in terms of the production of electrons
and photons (mostly from $\Pgpz$ decays), as well as hadrons (mostly $\Pgppm$) in the very forward direction
covered by the CASTOR calorimeter of the CMS experiment at the CERN LHC. CASTOR~\cite{Andreev:2010zzb}
covers the pseudorapidity region $-6.6<\eta<-5.2$,
and can distinguish between electromagnetic and hadronic energy depositions. CASTOR is only installed on the negative $z$-side of CMS, and hence has acceptance at negative pseudorapidities. Because of CASTOR's very
forward location, the data are sensitive to parton interactions at very small and large
fractional momenta in the proton, $x<10^{-4}$ and $x\to 1$.

\section{Experimental setup and Monte Carlo simulation}
The central feature of the CMS apparatus is a superconducting solenoid of 6\,m internal diameter,
providing a magnetic field of 3.8\unit{T}~\cite{Chatrchyan:2008zzk}. Within the field volume in the central region are a silicon pixel and strip tracker,
a lead tungstate crystal electromagnetic calorimeter, and a brass and scintillator hadron
calorimeter. Muons are measured in gas-ionization detectors embedded in the steel return yoke.

The central detectors of CMS are complemented by calorimeters in the
forward direction, which all rely on the detection of Cherenkov
photons produced when charged particles pass through their active quartz components. The ``hadron forward'' (HF)
calorimeters cover the pseudorapidity interval
$3.0<\abs{\eta}<5.2$ and use quartz fibers embedded in a steel
absorber. The CASTOR calorimeter is a sampling calorimeter composed of layers of fused
silica quartz plates and tungsten absorbers, segmented in 16
azimuthal towers, each with 14 longitudinal channels. The two
front channels have a combined depth of 20 radiation lengths and form the electromagnetic section of each tower. The remaining 12
channels constitute the hadronic section.
The full depth of a tower amounts to 10 hadronic interaction lengths.
A more detailed description of the CMS detector, together with a definition of the coordinate system used and all relevant kinematic variables, can be found in Ref.~\cite{Chatrchyan:2008zzk}. For triggering purposes, the Beam Pickup Timing for the eXperiment device was used~\cite{Khachatryan:2016bia}.

The corrections to the level of stable particles with $c\tau > 1\cm$ are determined by means of a
Monte Carlo (MC) simulation of the CMS apparatus based on \GEANTfour~\cite{Agostinelli2003250}, including
all known information about the CASTOR detector.

The data are compared to model predictions from \PYTHIA8~\cite{Sjostrand:2014zea} (version 8.212) with tune CUETP8M1~\cite{Khachatryan:2015pea},
which is based on measurements of the underlying event in p\={p} and pp collisions at $\sqrt{s}=1.96$ and 7\TeV, and
tune 4C~\cite{Corke:2010yf} combined with the MBR~\cite{Ciesielski:2012mc} model to describe diffractive processes.
The \PYTHIA8 CUETP8M1 tunes use the NNPDF2.3LO~\cite{Ball:2013hta} parton distribution functions (PDF), whereas tune 4C uses the CTEQ6L1 PDF~\cite{Lai:1999wy}.
Hadronic interaction event generators mostly developed for cosmic ray physics are also used:
\textsc{epos}~LHC~\cite{Pierog:2013ria} and its previous version \textsc{epos}~1.99~\cite{Pierog2009102}, \textsc{QGSJetII}~\cite{PhysRevD.83.014018} version
{II.3} and {II.4}, as well as \textsc{Sibyll}~2.1~\cite{PhysRevD.80.094003} and the recently released  \textsc{Sibyll}~2.3~\cite{Riehn:2015tbd}. The latest
versions of all these models are tuned to LHC data up to
$\sqrt{s}=8\TeV$, while the earlier versions are tuned to Tevatron results~\cite{dEnterria:2011twh}.

\section{Event selection and data analysis}

The present analysis is based on data that were recorded during the low luminosity LHC Run in 2015,
when CASTOR was operational and the CMS solenoid was off.
The data correspond to an integrated luminosity of
0.34\mubinv, with an average pp interaction probability
of 5\% per bunch crossing. Data were recorded with an unbiased
trigger requiring only the presence of two colliding bunches. Electronic
noise and beam-induced backgrounds are studied with data taken
without colliding bunches. Events are selected offline by requiring hadronic activity
in the HF calorimeters on either side of CMS. At least one
reconstructed calorimeter tower with energy larger than 5\GeV is
required. With these selection criteria the residual contribution of
electronic noise and beam background is well below 1\%.

Beam halo muons are used to determine the calibration of each CASTOR channel relative to the others.
This inter-calibration procedure cannot be applied to the last two longitudinal channels, which have detector noise levels
very close to the muon ionization peak and are not included in the dedicated halo-muon trigger.
These channels are therefore excluded from the analysis.

The response of CASTOR to pions and electrons was measured in a test beam in 2008~\cite{Gottlicher:2010zz}.
However, the configuration of CASTOR changed since then. Because of this, an independent method based
on 7\TeV collision data is used to determine the absolute energy scale calibration of CASTOR.
The average energy measured by the HF
calorimeters in the region $3<\abs{\eta}<5$ is fully
corrected to the particle level~\cite{Chatrchyan:2011wm} and extrapolated to the region
covered by CASTOR, using various hadronic interaction models.
The result of the extrapolation is used to calibrate CASTOR.
The detector response is found to be consistent with the test beam results.
Such a data-driven method facilitates the assignment of a
realistic uncertainty on the calorimeters energy scale.

\begin{figure}
\centering
\includegraphics[width=0.325\textwidth]{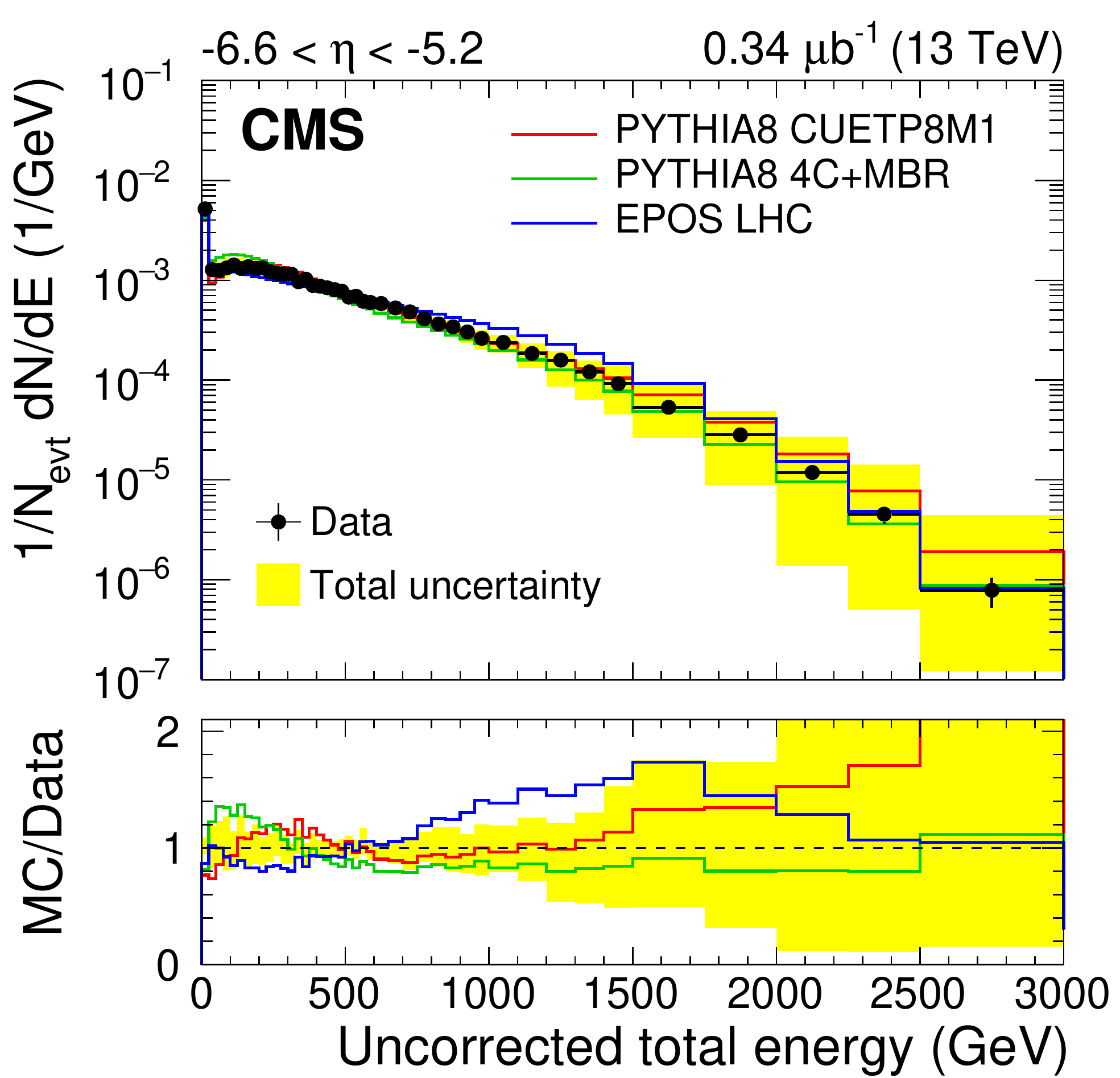}
\hfill
\includegraphics[width=0.325\textwidth]{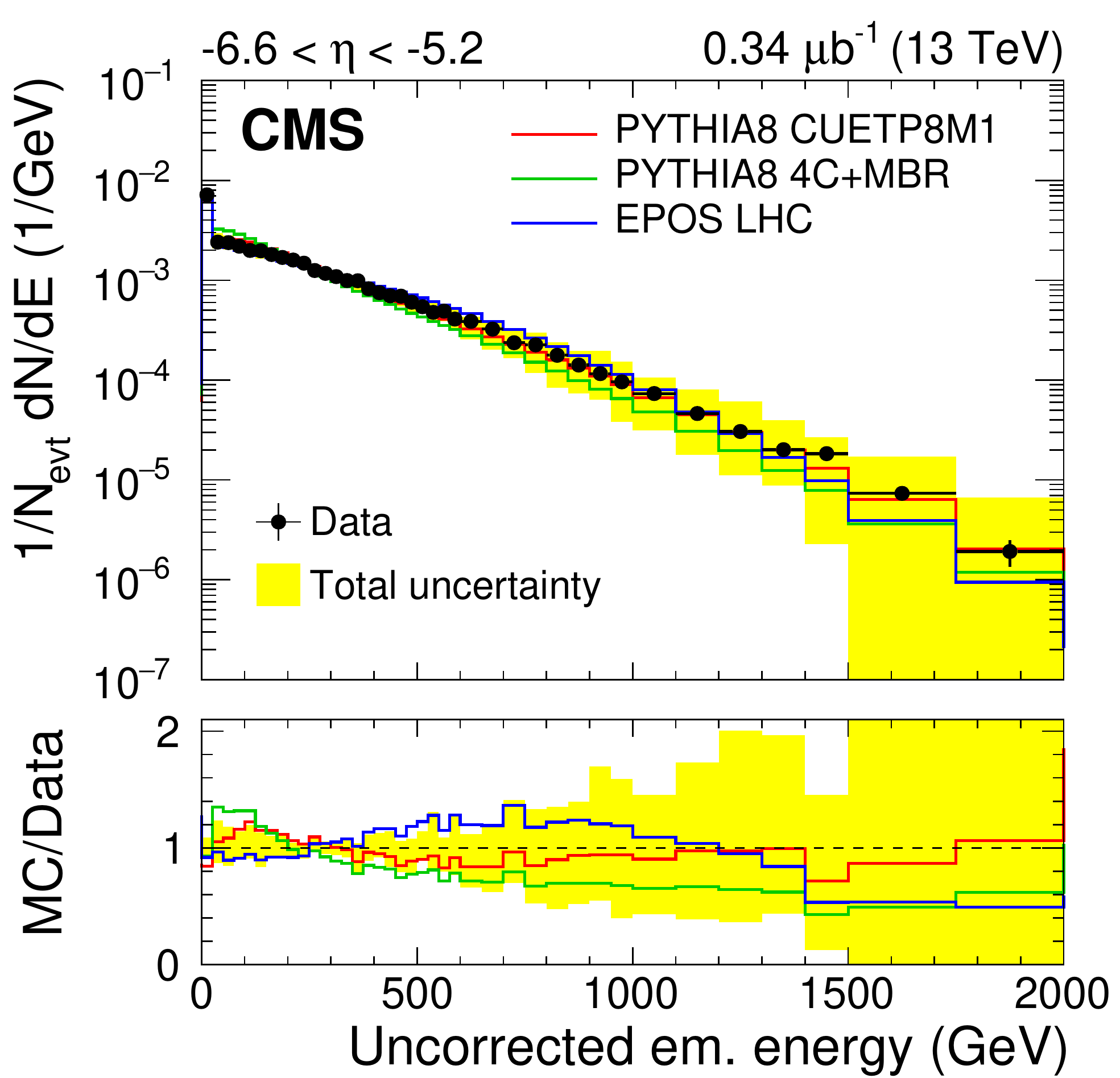}
\hfill
\includegraphics[width=0.325\textwidth]{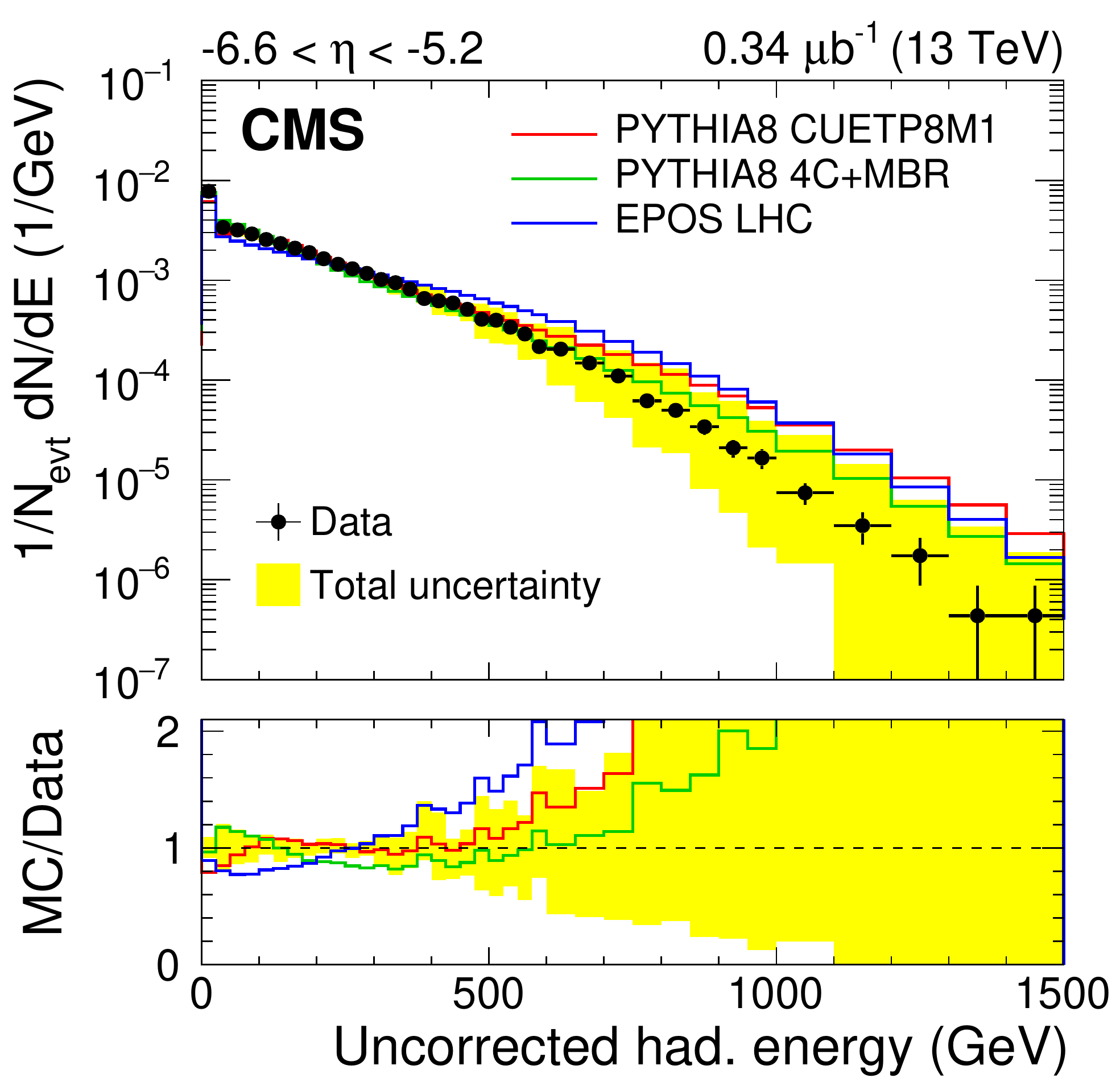}
\caption{
Spectra of the energy reconstructed in CASTOR, normalized to the number of events that pass the offline
event selection, compared to the detector-level predictions of various event generators.
The total energy spectrum is shown in the left panel, the electromagnetic in the middle, and the hadronic in the right.
Statistical (systematic) uncertainties are shown with error bars (yellow band, discussed in Section~\ref{sec:UncertaintiesResults}).}
\label{fig:CASTORRecoEnergySpectra}
\end{figure}

In order to reconstruct the total energy deposited in CASTOR, the energies of all calorimeter towers above the noise threshold
are summed up. This threshold is determined independently for every calorimeter tower and varies between 2 and 2.5\GeV .
The electromagnetic and hadronic contributions to the total energy can be determined by using the corresponding sections of CASTOR. The measured detector-level spectra are shown in Fig.~\ref{fig:CASTORRecoEnergySpectra}. Differences among model predictions are apparent.

The correction to the particle level is carried out through an unfolding
technique by means of the \textsc{RooUnfold} package~\cite{Adye:2011gm} with the
iterative algorithm proposed by D'Agostini~\cite{D'Agostini:1994zf}.
The procedure is stopped after 4 (6, 5) iterations for the total (electromagnetic, hadronic) spectrum, when the change with respect to the previous step is less than one percent.
Figure 2 shows the distributions of reconstructed energy in CASTOR as a function of the true energy at particle level for \PYTHIA8~CUETP8M1.
The fact that the slope of the correlation for the hadronic energy is not unity reflects the noncompensating nature of the calorimeter.

The event selection at particle level is based on the Lorentz-invariant fractional momentum loss of the proton, $\xi$.
All final-state particles are divided into two systems, $X$ and $Y$, based on their rapidity with respect to the pair of particles with the largest separation in rapidity. All particles on the negative side of this gap are assigned to the system $X$, while the particles on the positive side are assigned to the system $Y$~\cite{Chatrchyan:2013gfi}. The invariant masses, $M_\mathrm{X}$ and $M_\mathrm{Y}$, of each system are calculated using the four-momenta of the individual particles, and the variable $\xi$ is defined as the maximum of their squared ratios to the center-of-mass energy:
\begin{equation}
\begin{aligned}
\xi_\mathrm{X} &= M_\mathrm{X}^2 / s,\quad\xi_\mathrm{Y} = M_\mathrm{Y}^2 / s,\quad\text{and}\\
\xi &= \max(\xi_\mathrm{X},\xi_\mathrm{Y}).
\end{aligned}
\end{equation}

The variable $\xi$, commonly used to describe diffractive processes, is well defined for any pp final-state and provides a particularly economical particle-level description of the phase space acceptance at forward rapidities, independent of the underlying particle production mechanism.
Events with $\xi > 10^{-6}$ at particle level are selected, with an efficiency of about 97.3\% and a purity of about 99.5\% with respect to the detector-level event selection.
The total energy at particle level is calculated by summing up the energies of all particles, except muons and neutrinos, within the acceptance of CASTOR.
Muons and neutrinos are excluded since they do not deposit relevant energies in the detector.
For the electromagnetic spectrum, only electrons and photons are used; the latter are excluded for the hadronic energy spectrum. The decay photons of neutral pions constitute the dominant contribution to the electromagnetic spectrum.

\begin{figure}
\centering
\includegraphics[width=\textwidth]{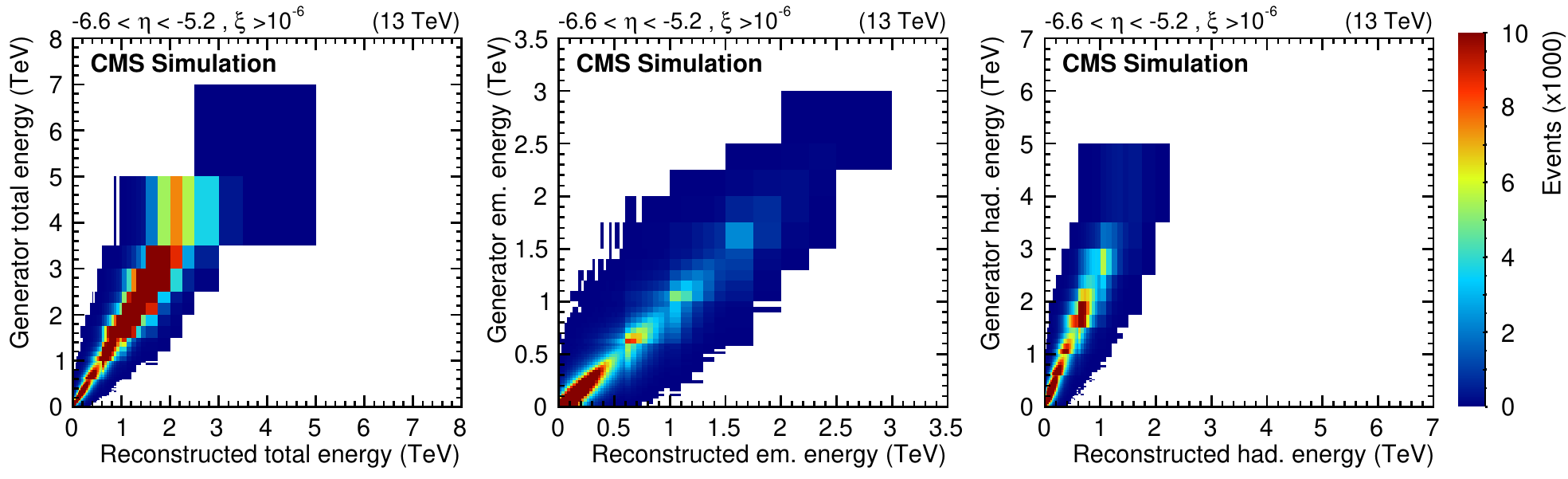}
\caption{Distributions of reconstructed energy as a function of the particle-level energy for \PYTHIA8~CUETP8M1 for the total (left), electromagnetic (middle), and hadronic (right) energy in CASTOR. The color indicates the number of events. The selection $\xi > 10^{-6}$ is explained in the text.}
\label{fig:ResponseMatrices}
\end{figure}

\section{Experimental uncertainties and results}
\label{sec:UncertaintiesResults}
The experimental uncertainties of the present results are mainly of systematic nature, with the CASTOR energy scale uncertainty being the most significant contribution.
In the data-driven calibration method, uncertainties on the energy scale arise from the HF energy scale uncertainty, the
extrapolation uncertainty, and the noncompensating calorimetric response of CASTOR. Furthermore, the energy measured by CASTOR depends on its exact location
with respect to the interaction point. This is because the energy flow $\rd E/\rd\eta$
rises sharply with $\eta$ in the very forward region. For the present data, the position of CASTOR is
known to within 1\mm, leading to a 7.5\% energy scale uncertainty. This is determined by means of Monte Carlo studies in which CASTOR is moved within the
measurement uncertainties. All contributions to the energy scale uncertainty add up to 17\% at detector level.

An additional systematic uncertainty comes from the inter-calibration of the channels with respect to each other.
This affects the separation of the electromagnetic and hadronic energies by up to 16\%.

The sensitivity of the result to the event selection based on activity in HF is quantified by varying by 10\% the 5\GeV selection threshold, which corresponds to the energy scale uncertainty of the HF calorimeters. The effect is below 6.2\% for the total energy, and less for the electromagnetic and hadronic energies.
Other sources of uncertainties, such as noise, beam background, or pileup are found to be negligible.

The detector-level spectra are varied within each of the above uncertainties, and then unfolded.
The spread of the unfolded spectra is then taken as a measure of the systematic uncertainty associated to each distribution.
Since the unfolding relies on Monte Carlo simulation, three models are used to unfold the detector-level spectra: \PYTHIA8~4C+MBR, \PYTHIA8~CUETP8M1, and \textsc{epos}~LHC.
The average of the resulting spectra is used as the nominal result and half their spread as an additional model-dependent systematic uncertainty.
This uncertainty is below 20\% for the total and electromagnetic spectra. The model dependence for the hadronic energy is higher and reaches 63\% in some energy bins. These uncertainties increase with energy.
The luminosity recorded by CMS is determined with a precision of 2.3\% for data taken with full magnetic field~\cite{CMS-PAS-LUM-15-001}. The luminosity at zero magnetic field can be recalibrated by comparing full and zero field data directly; the corresponding uncertainty is 2.6\%.

All contributions to the systematic uncertainties are added in quadrature.
Example values are given for two bins of total, hadronic, and electromagnetic energies in Table~\ref{tab:SysUncertainty}.
The statistical uncertainties are calculated with the full covariance matrices of the unfolded results and therefore include correlations induced by the unfolding procedure. The full covariance matrices are available in the \textsc{HEPData} record of this paper; the differential cross section measurements with uncertainties are also tabulated therein~\cite{HEPData}. In the figures, the total uncertainties are shown as yellow bands; they include the statistical uncertainties, which in most bins are not visible.
The uncertainty assigned to the model dependence of the unfolding procedure is shown as an orange band.

\begin{figure}
\centering
\includegraphics[width=0.48\textwidth]{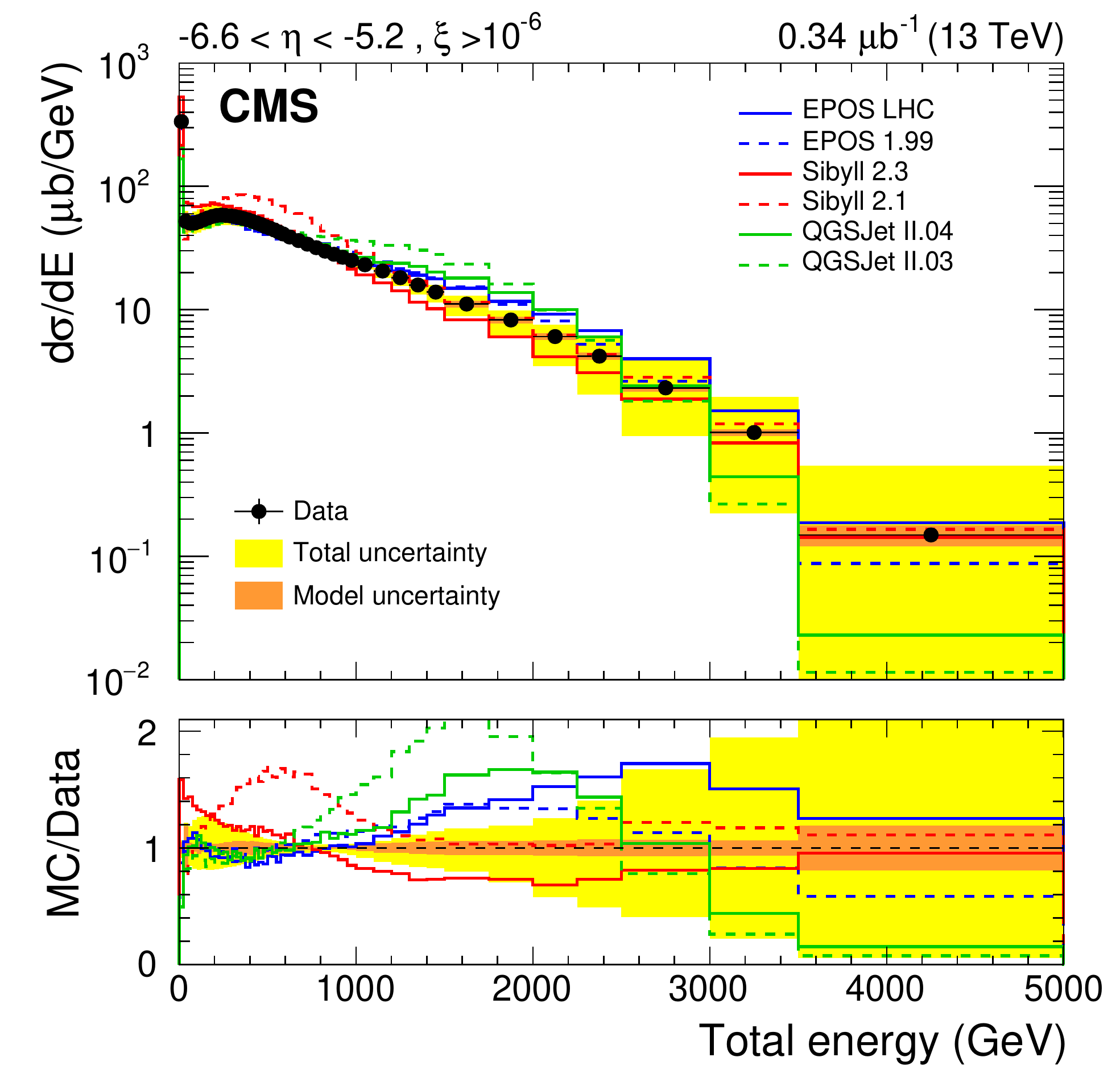}
\hfill
\includegraphics[width=0.48\textwidth]{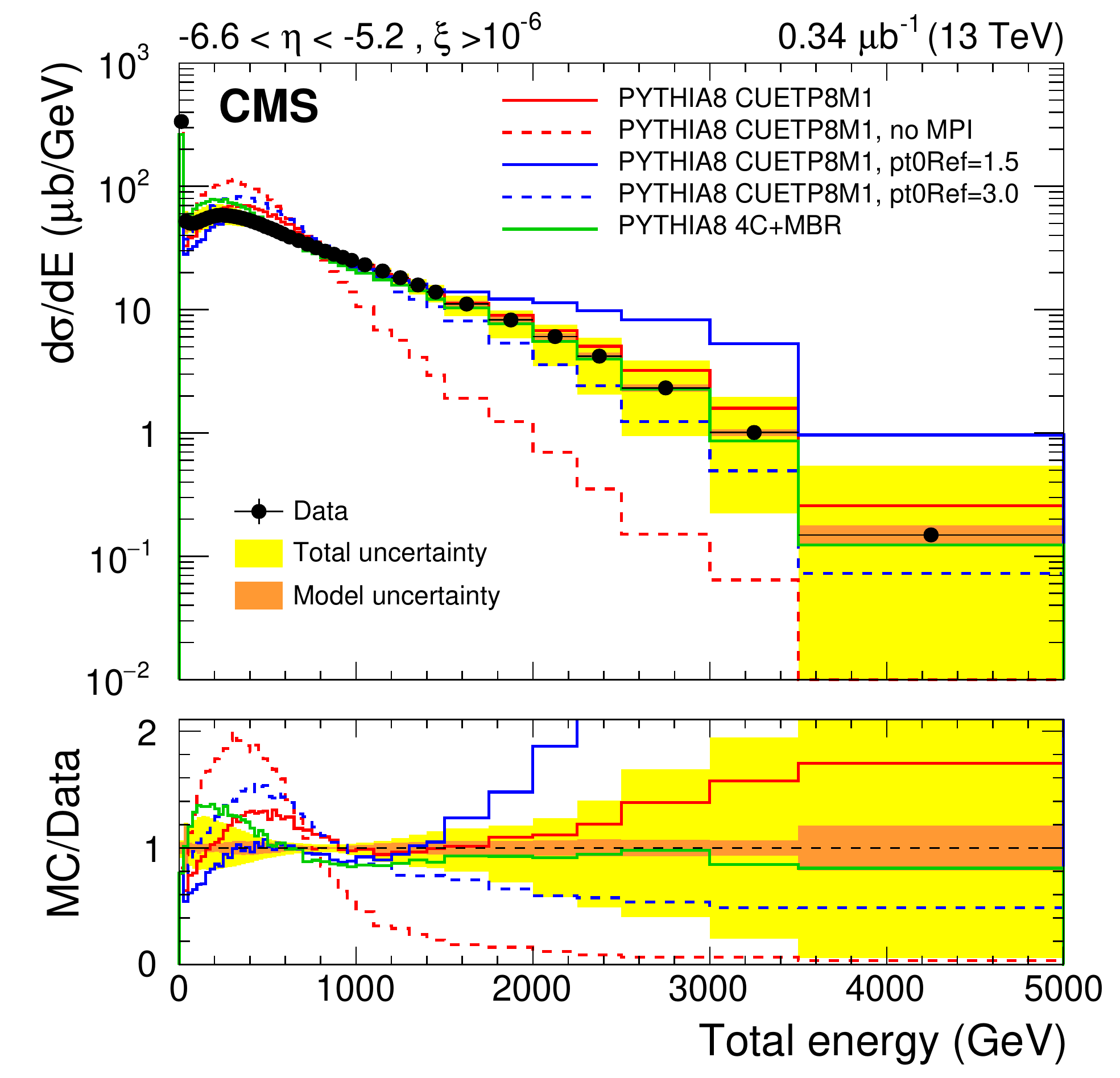}
\caption{Differential cross section as a function of the total energy in the
region $-6.6<\eta < -5.2$ for events with $\xi>10^{-6}$. The left panel shows
the data compared to MC event generators mostly developed for cosmic ray induced
air showers, and the right panel to different \PYTHIA8 tunes.}
\label{fig:Result_all}
\end{figure}

\begin{figure}
\centering
\includegraphics[width=0.48\textwidth]{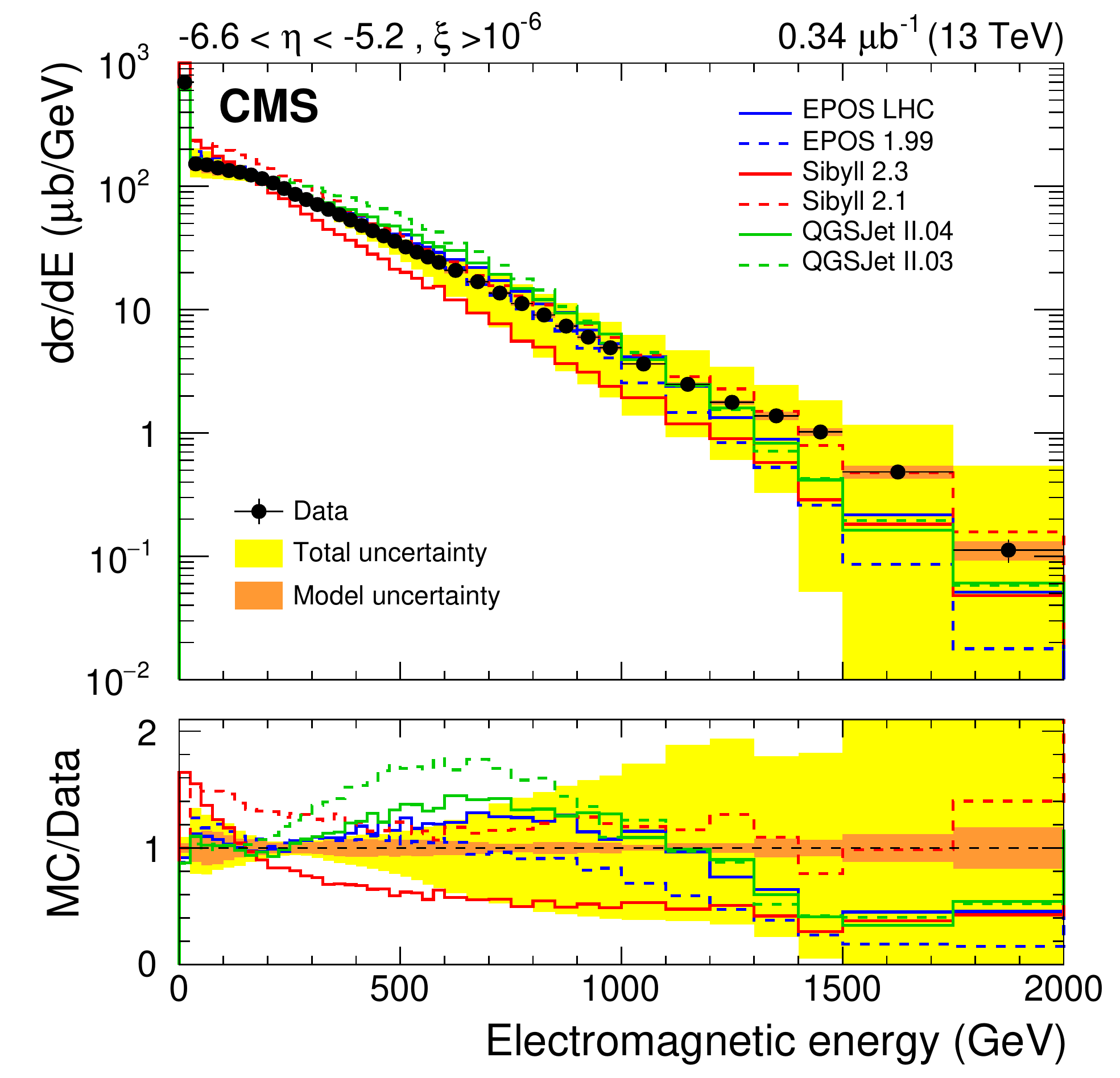}
\hfill
\includegraphics[width=0.48\textwidth]{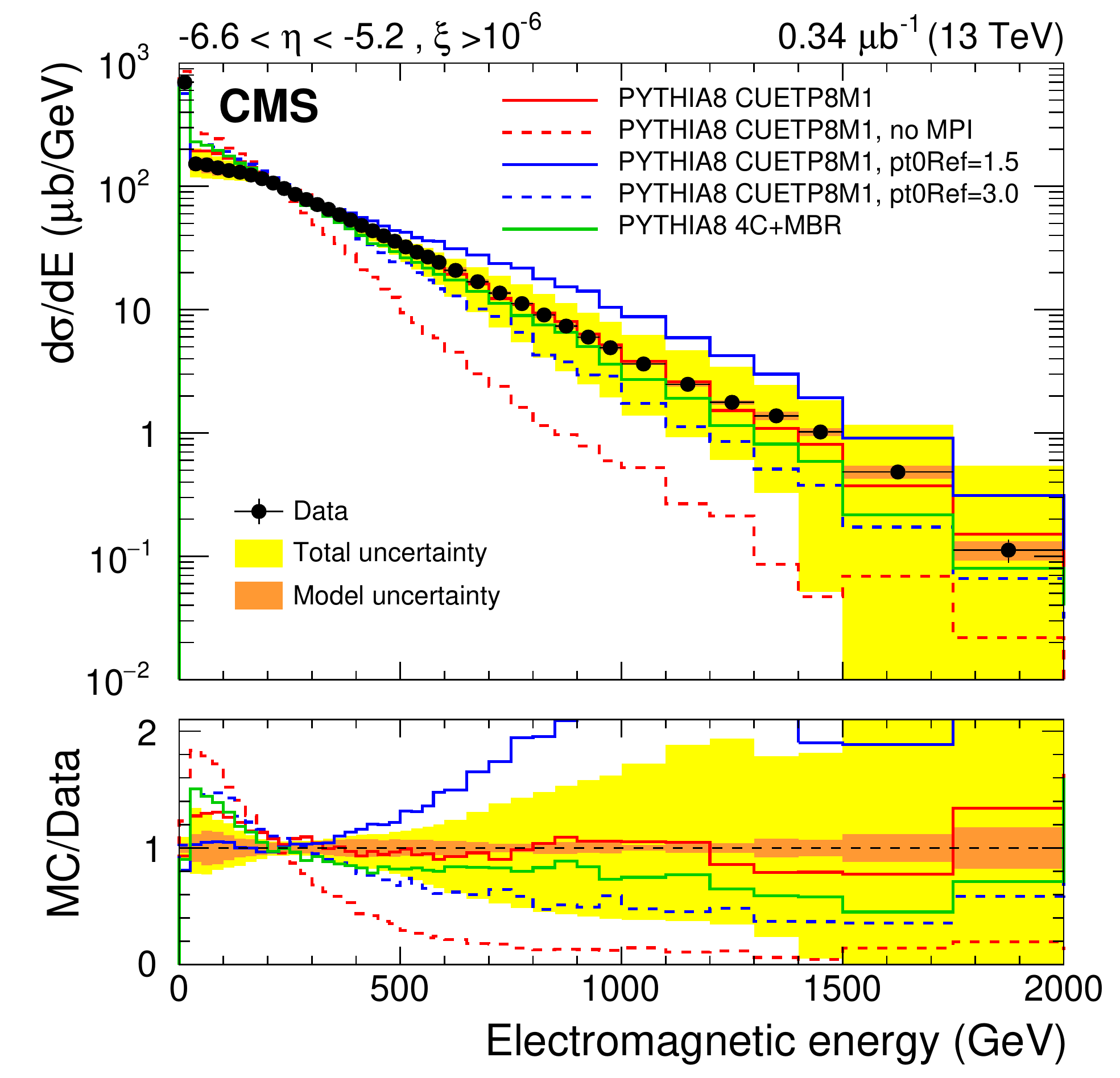}
\caption{Differential cross section as a function of the electromagnetic energy in the
region $-6.6<\eta < -5.2$ for events with $\xi>10^{-6}$. The left panel shows
the data compared to MC event generators mostly developed for cosmic ray induced
air showers, and the right panel to different \PYTHIA8 tunes.}
\label{fig:Result_em}
\end{figure}

\begin{figure}
\centering
\includegraphics[width=0.48\textwidth]{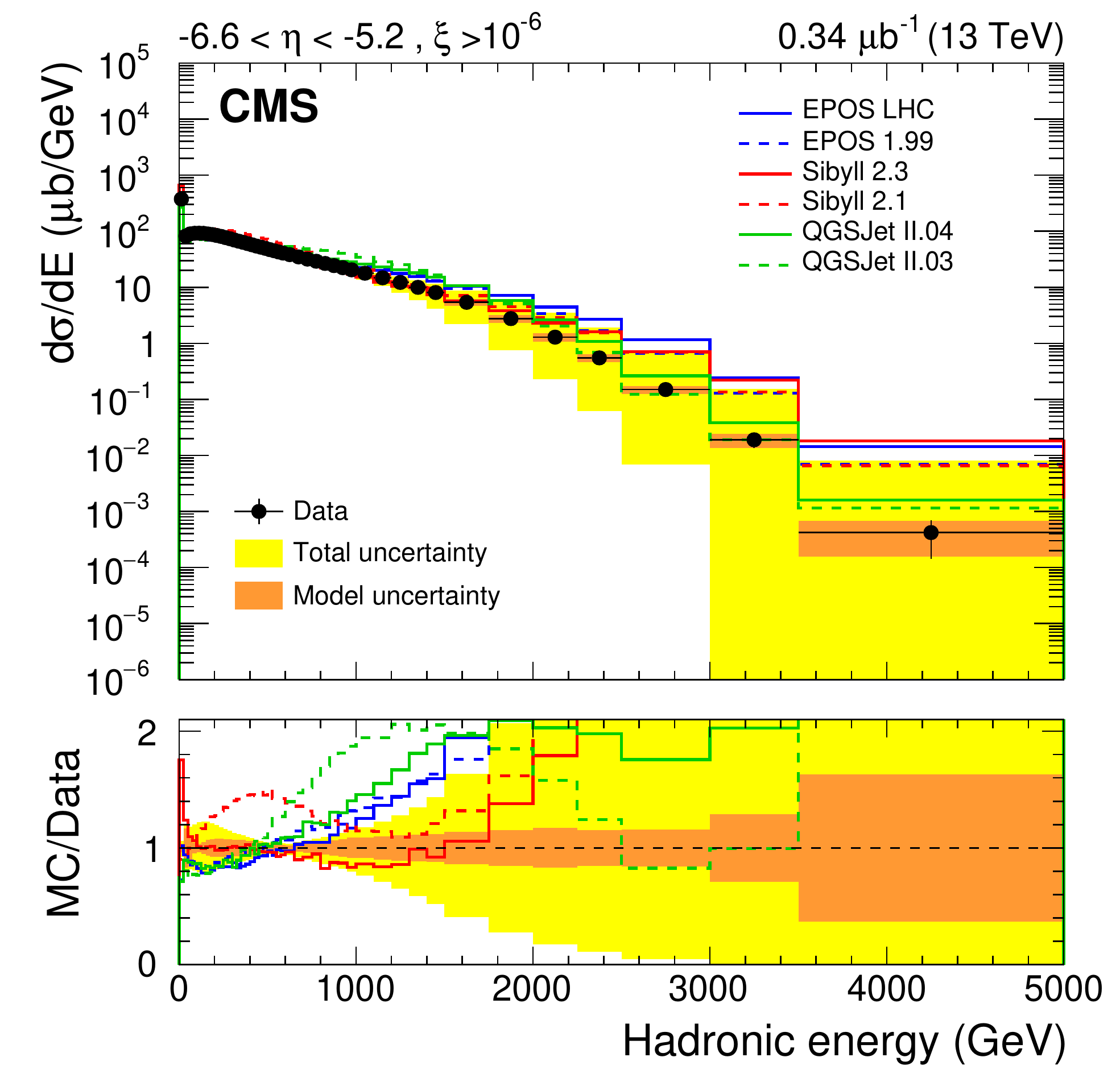}
\hfill
\includegraphics[width=0.48\textwidth]{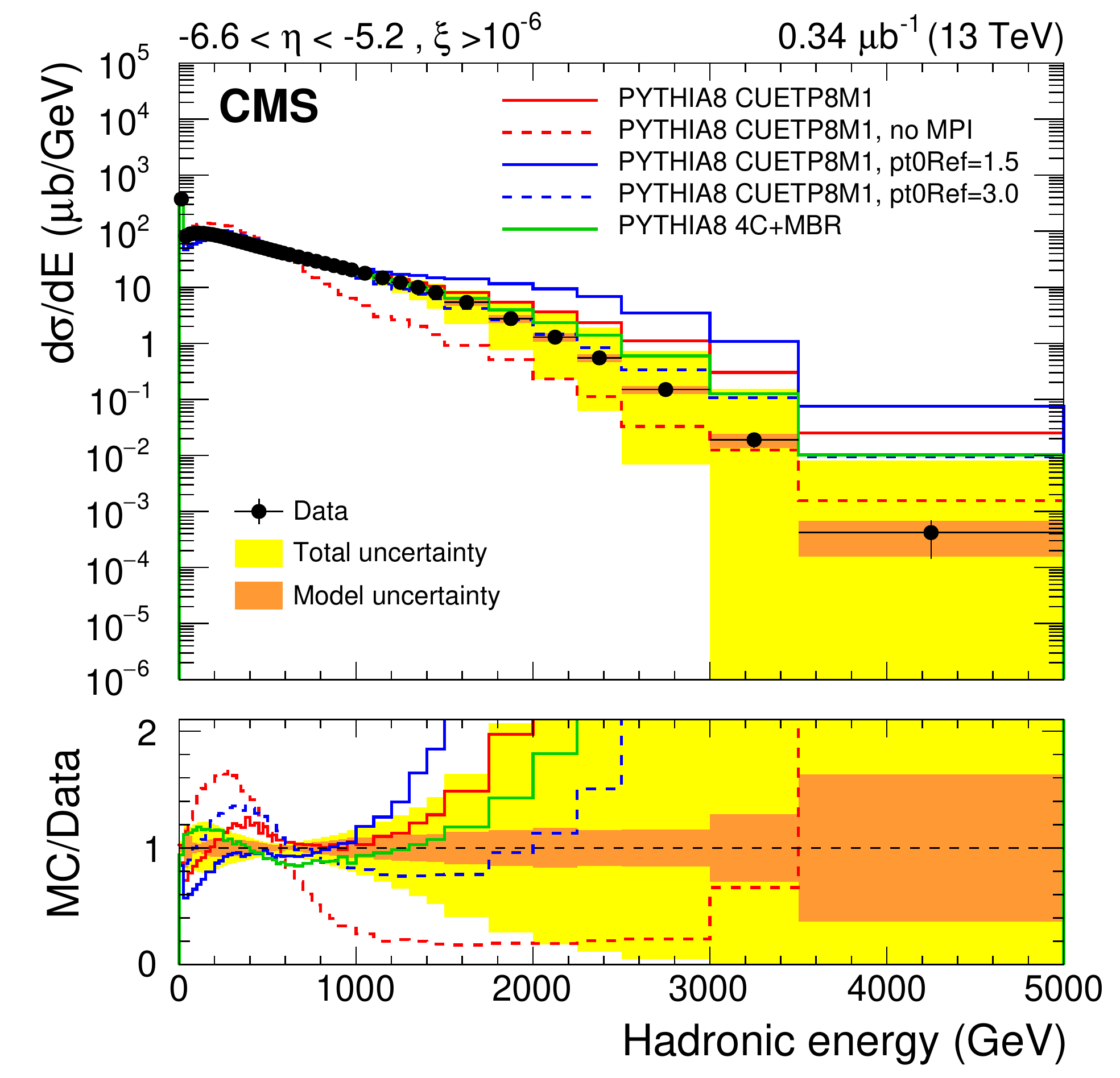}
\caption{Differential cross section as a function of the hadronic energy in the
region $-6.6<\eta < -5.2$ for events with $\xi>10^{-6}$. The left panel shows
the data compared to MC event generators mostly developed for cosmic ray induced
air showers, and the right panel to different \PYTHIA8 tunes.}
\label{fig:Result_had}
\end{figure}

The total, electromagnetic, and hadronic energy spectra are measured in the region $-6.6<\eta < -5.2$ and corrected to the particle level for $\xi>10^{-6}$. They are shown in Figs.~\ref{fig:Result_all}--\ref{fig:Result_had} and compared to the predictions of \textsc{epos}, \textsc{QGSJetII} and \textsc{Sibyll} (left plots) and various \PYTHIA8 tunes (right plots). All spectra feature a sharp peak at zero reflecting the presence of diffractive events with forward rapidity gap(s). The total and hadronic energy spectra exhibit peaks at about 300 and 100\GeV respectively, followed by a long tail towards higher energies.
Such a peak is not observed in the electromagnetic spectrum.

In Fig. \ref{fig:Result_all}, the distribution of the total energy is shown. Different parts of the spectrum are reproduced by different models.
None of the models reproduce all features of the data, but the bump at about 300\GeV is visible in all of them.
The spectrum is best described by \textsc{epos}~LHC and \textsc{QGSJetII.4}. The \PYTHIA8 tunes tend to overestimate the contribution of the soft part of the spectrum and so does \textsc{Sibyll}~2.3. The high energy tail is well described by \PYTHIA8 and \textsc{Sibyll}, whereas \textsc{epos}~LHC and \textsc{QGSJetII.4} overestimate the region between 1 and 2.5\TeV . The predictions are also very sensitive to the scaling parameter
$p_{\mathrm{T,0}}^{\text{ref}}$ of \PYTHIA8, which parameterizes the dampening of the partonic cross sections for $\pt\to 0$ at the nominal
reference energy of 7\TeV ~\cite{Sjostrand:2014zea}. Large deviations from
the CUETP8M1 $p_{\mathrm{T,0}}^{\text{ref}}$ default value of 2.4024\GeV modify the simulation of
MPI and lead to predictions inconsistent with the data; the PDF choice (NNPDF2.3 in CUETP8M1 versus CTEQ6L1 in 4C) and/or the diffraction model used (MBR or default \PYTHIA8 model) also play a role in some regions of phase space.

The electromagnetic spectrum is shown in Fig. \ref{fig:Result_em}; it is relatively well described by most of the models within uncertainties.
Only \PYTHIA8~4C+MBR and \textsc{Sibyll}~2.3 do
not correctly model the shape of the soft part of the spectrum up to
about 500\GeV . The comparison of the data to the predictions of various \PYTHIA8 tunes indicates that the electromagnetic energy distribution
is also very sensitive to the underlying modeling of MPI.

Figure~\ref{fig:Result_had} shows the hadronic energy distribution. While \textsc{epos}~LHC and \textsc{QGSJetII} perform well at lower energies, they predict too large a cross section in the range of 600 to 1800\GeV .
This feature is also observed in the total energy spectrum, suggesting that the excess
originates from the production of hadrons.
\textsc{Sibyll}~2.3 reproduces the slope of the spectrum over a larger energy range, but significantly overestimates the cross section at very low energy, while \textsc{Sibyll}~2.1 shows a large excess at around 500\GeV , similar to that observed in the total energy spectrum.

{
\renewcommand{\arraystretch}{1.2}
\begin{table}
\centering
\topcaption{Uncertainties on the differential cross sections at a few selected values of the total, electromagnetic, and hadronic energies.}
\label{tab:SysUncertainty}
\begin{tabular}{rcc|cc|cc}
                     & \multicolumn{2}{c}{Total}                                     & \multicolumn{2}{c}{Electromagnetic}
                     & \multicolumn{2}{c}{Hadronic} \\
                     & 300\GeV                       & 3000\GeV                      & 300\GeV                         & 1200\GeV
                     & 300\GeV                       & 2000\GeV    \\ \hline
Energy Scale     & $\substack{+17\,\%\\-14\,\%}$ & $\substack{+94\,\%\\-77\,\%}$ & $\substack{+5.9\,\%\\-2.1\,\%}$ & $\substack{+93\,\%\\-65\,\%}$
                & $\substack{+11\,\%\\-10\,\%}$ & $\substack{+169\,\%\\-80\,\%}$ \\
Unfolding           & $\substack{\pm5.8\%}$         & $\substack{\pm6.4\%}$         & $\substack{\pm5.2\%}$           & $\substack{\pm4.1\%}$
                & $\substack{\pm6.9\%}$         & $\substack{\pm17\%}$ \\
Event selection & $\substack{\pm0.5\%}$         & $\substack{< 0.01\%}$         & $\substack{\pm0.14\%}$          & $\substack{< 0.01\%}$
                & $\substack{\pm0.06\%}$         & $\substack{< 0.01\%}$ \\ \hline
Luminosity      & \multicolumn{6}{c}{$\substack{\pm2.6\%}$} \\ \hline
Statistical     & $\substack{\pm1.2\%}$         & $\substack{\pm4.3\%}$         & $\substack{\pm1.5\%}$           & $\substack{\pm5.9\%}$
                     & $\substack{\pm1.0\%}$         & $\substack{\pm4.2\%}$ \\ \hline
\end{tabular}
\end{table}
}


\section{Summary}

The electromagnetic, hadronic, and total energy spectra of particles produced at very forward
pseudorapidities ($-6.6<\eta < -5.2$) have been measured with the CASTOR calorimeter of the
CMS experiment in proton-proton collisions at a center-of-mass energy of 13\TeV . The experimental distributions, fully corrected for detector effects, are compared to the predictions of various Monte Carlo event generators commonly used in high energy cosmic ray physics (\textsc{EPOS}, \textsc{QGSJetII}, and \textsc{Sibyll}), and those of different tunes of \PYTHIA8. None of the generators considered describe all features seen in the data.

The present measurements are particularly sensitive to the modeling of multiparton interactions (MPI) that dominate particle production
in the underlying event at forward rapidities in pp collisions. \PYTHIA8 CUETP8M1 without MPI is ruled out by the data, which exhibit much harder spectra than predicted by the model. The shape of the spectra are significantly influenced by the
MPI-related settings in \PYTHIA8. The present results can therefore contribute to improvements in future Monte Carlo parameter tunes.

Event generators developed for modeling high energy cosmic ray air showers, tuned to LHC measurements at 0.9, 7, and 8\TeV , agree
better with the present data than those tuned to Tevatron results alone. This is
especially true for \textsc{QGSJetII} and \textsc{Sibyll}.
However, all these models underestimate the muon production rate in extensive air showers because of their inaccurate
description of the hadronic shower component~\cite{Aab:2016hkv}. The present results provide new constraints for
improving the modeling of hadron production in event generators commonly used in high energy particle and cosmic ray physics.

\begin{acknowledgments}
We congratulate our colleagues in the CERN accelerator departments for the excellent performance of the LHC and thank the technical and administrative staffs at CERN and at other CMS institutes for their contributions to the success of the CMS effort. In addition, we gratefully acknowledge the computing centers and personnel of the Worldwide LHC Computing Grid for delivering so effectively the computing infrastructure essential to our analyses. Finally, we acknowledge the enduring support for the construction and operation of the LHC and the CMS detector provided by the following funding agencies: BMWFW and FWF (Austria); FNRS and FWO (Belgium); CNPq, CAPES, FAPERJ, and FAPESP (Brazil); MES (Bulgaria); CERN; CAS, MoST, and NSFC (China); COLCIENCIAS (Colombia); MSES and CSF (Croatia); RPF (Cyprus); SENESCYT (Ecuador); MoER, ERC IUT, and ERDF (Estonia); Academy of Finland, MEC, and HIP (Finland); CEA and CNRS/IN2P3 (France); BMBF, DFG, and HGF (Germany); GSRT (Greece); OTKA and NIH (Hungary); DAE and DST (India); IPM (Iran); SFI (Ireland); INFN (Italy); MSIP and NRF (Republic of Korea); LAS (Lithuania); MOE and UM (Malaysia); BUAP, CINVESTAV, CONACYT, LNS, SEP, and UASLP-FAI (Mexico); MBIE (New Zealand); PAEC (Pakistan); MSHE and NSC (Poland); FCT (Portugal); JINR (Dubna); MON, RosAtom, RAS, RFBR and RAEP (Russia); MESTD (Serbia); SEIDI, CPAN, PCTI and FEDER (Spain); Swiss Funding Agencies (Switzerland); MST (Taipei); ThEPCenter, IPST, STAR, and NSTDA (Thailand); TUBITAK and TAEK (Turkey); NASU and SFFR (Ukraine); STFC (United Kingdom); DOE and NSF (USA).

\hyphenation{Rachada-pisek} Individuals have received support from the Marie-Curie program and the European Research Council and EPLANET (European Union); the Leventis Foundation; the A. P. Sloan Foundation; the Alexander von Humboldt Foundation; the Belgian Federal Science Policy Office; the Fonds pour la Formation \`a la Recherche dans l'Industrie et dans l'Agriculture (FRIA-Belgium); the Agentschap voor Innovatie door Wetenschap en Technologie (IWT-Belgium); the Ministry of Education, Youth and Sports (MEYS) of the Czech Republic; the Council of Science and Industrial Research, India; the HOMING PLUS program of the Foundation for Polish Science, cofinanced from European Union, Regional Development Fund, the Mobility Plus program of the Ministry of Science and Higher Education, the National Science Center (Poland), contracts Harmonia 2014/14/M/ST2/00428, Opus 2014/13/B/ST2/02543, 2014/15/B/ST2/03998, and 2015/19/B/ST2/02861, Sonata-bis 2012/07/E/ST2/01406; the National Priorities Research Program by Qatar National Research Fund; the Programa Clar\'in-COFUND del Principado de Asturias; the Thalis and Aristeia programs cofinanced by EU-ESF and the Greek NSRF; the Rachadapisek Sompot Fund for Postdoctoral Fellowship, Chulalongkorn University and the Chulalongkorn Academic into Its 2nd Century Project Advancement Project (Thailand); and the Welch Foundation, contract C-1845.
\end{acknowledgments}
\bibliography{auto_generated}
\cleardoublepage \appendix\section{The CMS Collaboration \label{app:collab}}\begin{sloppypar}\hyphenpenalty=5000\widowpenalty=500\clubpenalty=5000\textbf{Yerevan Physics Institute,  Yerevan,  Armenia}\\*[0pt]
A.M.~Sirunyan, A.~Tumasyan
\vskip\cmsinstskip
\textbf{Institut f\"{u}r Hochenergiephysik,  Wien,  Austria}\\*[0pt]
W.~Adam, E.~Asilar, T.~Bergauer, J.~Brandstetter, E.~Brondolin, M.~Dragicevic, J.~Er\"{o}, M.~Flechl, M.~Friedl, R.~Fr\"{u}hwirth\cmsAuthorMark{1}, V.M.~Ghete, C.~Hartl, N.~H\"{o}rmann, J.~Hrubec, M.~Jeitler\cmsAuthorMark{1}, A.~K\"{o}nig, I.~Kr\"{a}tschmer, D.~Liko, T.~Matsushita, I.~Mikulec, D.~Rabady, N.~Rad, B.~Rahbaran, H.~Rohringer, J.~Schieck\cmsAuthorMark{1}, J.~Strauss, W.~Waltenberger, C.-E.~Wulz\cmsAuthorMark{1}
% ...
\vskip\cmsinstskip
% ...
1:~~Also at Vienna University of Technology, Vienna, Austria\\
% ...

\end{sloppypar}
\end{document}